\documentclass[aps,pre,twocolumn,groupedaddress,showpacs,showkeys]{revtex4}

\usepackage{graphicx}
\usepackage{dcolumn}
\usepackage{bm}
\usepackage{amssymb}
\usepackage{amsmath}
\begin{document}

\title{Generalized synchronization in coupled Ginzburg-Landau equations
and mechanisms of its arising\footnote{This
  paper was published in Physical Review E (Statistical,
  Nonlinear, and Soft Matter Physics) Volume 72, No 3 (2005)
  037201}}
\author{Alexander~E.~Hramov}
 \email{aeh@cas.ssu.runnet.ru}
\author{Alexey~A.~Koronovskii}
 \email{alkor@cas.ssu.runnet.ru}
\author{Pavel V. Popov}
\affiliation{Faculty of Nonlinear Processes, Saratov State
University, Astrakhanskaya, 83, Saratov, 410012, Russia}

\begin{abstract}
Generalized chaotic synchronization regime is observed in the
unidirectionally coupled one-dimensional Ginzburg-Landau
equations. The mechanism resulting in the generalized
synchronization regime arising in the coupled spatially extended
chaotic systems demonstrating spatiotemporal chaotic oscillations
has been described. The cause of the generalized synchronization
occurrence is studied with the help of the modified
Ginzburg-Landau equation with additional dissipation.
\end{abstract}

\date{\today}


\maketitle

Chaotic synchronization is one of the fundamental phenomena
actively studied recently \cite{Pikovsky:2002_SynhroBook,
Boccaletti:2002_ChaosSynchro}, having both important theoretical
and applied significance (e.g., used for information
trans\-mis\-sion by means of deterministic chaotic signals
\cite{Murali:1993_SignalTransmission, Chua:1997_Criptography}, in
biological \cite{Elson:1998_NeronSynchro} and physiological
\cite{Prokhorov:2003_HumanSynchroPRE} tasks, for controlling of
lasers \cite{Roy:1992ControlChaos,
Fischer:2000_ChaosCommunication} and microwave systems
\cite{Hramov:2005_Chaos_BWO} etc.).
Recently, several types of chaotic synchronization have been
observed in coupled nonlinear oscillators. These are the phase
synchronization \cite{Rosenblum:1996_PhaseSynchro}, generalized
synch\-roni\-za\-tion \cite{Rulkov:1995_GeneralSynchro}, lag
synch\-roni\-za\-tion \cite{Rosenblum:1997_LagSynchro},
intermittent lag \cite{Boccaletti:2000_IntermitLagSynchro} and
intermittent generalized \cite{Hramov:2005_IGS_EuroPhysicsLetters}
synchronization  behaviour, complete synchronization
\cite{Pecora:1990_ChaosSynchro}.

All synchronization types mentioned above are associated with each
other (see, for detail, \cite{Boccaletti:2001_UnifingSynchro,
Brown:2000_ChaosSynchro, Hramov:2004_Chaos,
Aeh:2005_SpectralComponents}), but the relationship between them
has not been completely clarified yet. In particular, in our works
\cite{Hramov:2004_Chaos, Aeh:2005_SpectralComponents,
Aeh:2005_TSS:PhysicaD} it was shown that the phase, generalized,
lag, and complete synchronization are closely connected with each
other and, as a matter of fact, they are different manifestations
of one type of synchronous oscillation behavior of coupled chaotic
oscillators called the time-scale synchronization. For each type
of synchronization there are their own ways to detect the
synchronized behavior of coupled chaotic oscillators.

In the last decade synchronization of spatially extended systems
demonstrating spatio\-tempo\-ral chaos has attracted much
interest. The possibility of the complete synchronization and
phase synchronization of spatially extended systems such as
coupled Ginzburg-Landau equations
\cite{Boccaletti:2002_ChaosSynchro, Boccaletti:1999_GLE_Synchro,
bragard:036219}, coupled Kuramoto-Sivashinsky equations
\cite{Tasev:2000_Kur_Siv_Synchro}, arrays of coupled oscillators
\cite{Boccaletti:1999_ControllingChaos}, and coupled map lattices
\cite{Boccaletti:2002_ChaosSynchro} has been demonstrated
recently. In particular, the experimental phase synchronization
has been observed for a plasma discharge tube in
work~\cite{Rosa:2000_PlasmaDischarge}. In our work
\cite{Hramov:2005_Chaos_BWO} we have shown that the time scale
synchronization takes place in unidirectionally coupled spatially
extended electron--wave systems.

One of the interesting and intricate types of the synchronous
behavior of unidirectionally coupled chaotic oscillators is the
generalized synchronization \cite{Rulkov:1995_GeneralSynchro}. The
presence of the generalized synchronization between the response
$\mathbf{x}_{r}(t)$ and drive $\mathbf{x}_{d}(t)$ chaotic systems
means that there is a functional relation
${\mathbf{x}_r(t)=\mathbf{F}[\mathbf{x}_d(t)]}$ between system
states after the transient is finished.
This functional relation $\mathbf{F}[\cdot]$ may be smooth or
fractal. According to the properties of this relation, the
generalized synchronization may be divided into the strong
synchronization and week synchronization,
respectively~\cite{Pyragas:1996_WeakAndStrongSynchro}. There are
several methods to detect the presence of the generalized
synchronization between chaotic oscillators, such as the auxiliary
system approach~\cite{Rulkov:1996_AuxiliarySystem} or the method
of calculating the conditional Lyapunov exponents
\cite{Pyragas:1996_WeakAndStrongSynchro}.

In this work we have used the auxiliary system approach proposed
firstly in~\cite{Rulkov:1996_AuxiliarySystem}. We consider the
dynamics of the drive $\mathbf{x}_d(t)$  and response
$\mathbf{x}_r(t)$ systems. At the same time we also consider the
dynamics of the auxiliary system $\mathbf{x}_a(t)$ which is
identical to the response system $\mathbf{x}_r(t)$ but starts with
the other initial conditions, i.e., $\mathbf{x}_r(t_0)\neq
\mathbf{x}_a(t_0)$. In the absence of the generalized
synchronization between the drive $\mathbf{x}_d(t)$ and response
$\mathbf{x}_r(t)$ systems, the phase trajectories of the response
$\mathbf{x}_r(t)$ and auxiliary $\mathbf{x}_a(t)$ systems share
the same chaotic attractor but are unrelated. In the case of the
generalized synchronization the behavior of the response
$\mathbf{x}_r(t)$ and auxiliary $\mathbf{x}_a(t)$ systems becomes
identical after the transient dies out (it may take much time
~\cite{Hramov:2005_IGS_EuroPhysicsLetters}) due to the generalized
synchronization relations
$\mathbf{x}_r(t)=\mathbf{F}[\mathbf{x}_d(t)]$ and
$\mathbf{x}_a(t)=\mathbf{F}[\mathbf{x}_d(t)]$. Obviously, in the
case of the generalized synchronization the condition
$\mathbf{x}_d(t)=\mathbf{x}_r(t)$ should be satisfied and the
identity of the response and auxiliary systems is a simpler
criterion to test the presence of the generalized synchronization
rather than finding the unknown functional relationship
$\mathbf{F}[\cdot]$.

Note, that the generalized synchronization has been studied in
detail only for the chaotic systems with few degrees of freedom
and for the discrete maps \cite{Rulkov:1995_GeneralSynchro,
Rulkov:1996_AuxiliarySystem, Pyragas:1996_WeakAndStrongSynchro}.
In particular, in work \cite{Aeh:2005_GS:ModifiedSystem} we have
shown that the behavior of the response chaotic system in the
regime of the generalized synchronization is equal to the dynamics
of the modified system (with the additional dissipation) under the
external chaotic force. However, the generalized synchronization
of the spatially extended chaotic systems has not been studied in
detail. Here  we note only work \cite{Parmananda:1997_GS} in which
the occurrence of the generalized synchronization in the spatially
extended model describing a chemical reaction has been found. The
mechanism of the establishment of the generalized synchronization
in the spatially extended chaotic systems is also unclear.

In this paper we study numerically the generalized synchronization
of the unidirectionally coupled complex Ginzburg-Landau equations
(CGLE's). The Ginzburg-Landau equation (GLE) is a fundamental
model for the pattern formation and turbulence description. This
equation is used frequently to describe many different phenomena
in laser physics \cite{Coullet:1989_Ginzburg-Landau}, chemical
turbulence \cite{Kuramoto:1981_Ginzburg-Landau}, fluid dynamics
\cite{Kolodner:1995_Ginzburg-Landau}, bluff body wakes
\cite{Leweke:1994_Ginzburg-Landau}, etc. (see also
\cite{Bragard:2000_SPACEEXTENDEDCHAOSSYNCHRO}).

Let us consider two one-dimensional CGLE's
\cite{Boccaletti:1999_GLE_Synchro, Junge:2000_Ginzburg-Landau,
Bragard:2000_SPACEEXTENDEDCHAOSSYNCHRO} coupled unidirectionally
\begin{equation}
\frac{\partial v}{\partial t}=
v-(1-i\alpha_d)|v|^2v+(1+i\beta_d)\Delta v, \quad v\in[0,L],
\label{eq:drive}
\end{equation}
\begin{equation}
\frac{\partial u}{\partial t}=
u-(1-i\alpha_r)|u|^2u+(1+i\beta_r)\Delta u + \varepsilon (v-u),
\quad u\in[0,L] \label{eq:response}
\end{equation}
with periodical boundary conditions. Equation (\ref{eq:drive})
describes the drive system and equation (\ref{eq:response})
corresponds to the response one. In our investigation the
parameters of the drive systems are chosen as $\alpha_d=1.5$,
$\beta_d=1.5$. To study the generalized synchronization of the
nonidentical systems we have chosen the different values of
control parameters ($\alpha_r=4.0$ and $\beta_r=4.0$) for the
response system (\ref{eq:response}). The choice of such values of
the control parameters results in the autonomous systems being in
the spatiotemporal chaotic regime. Parameter $\varepsilon$
determines the strength of the unidirectionally dissipative
coupling between the response and drive systems, the interaction
of them being in each point of space. For $\varepsilon=0$, Eqs.
(\ref{eq:drive}) and (\ref{eq:response}) describe two uncoupled
complex fields $u(x,t)$, $v(x,t)$, each of them obeying an
autonomous GLE.

All calculations were performed for a fixed system length
$L=40\pi$ and random initial conditions. The numerical code was
based on a semi-implicit scheme in time with finite differences in
space. In all simulations we used a time step $\Delta t=0.0002$
for the integration and a space discretization $\Delta x=L/1024$
(1024 mesh points).

\begin{figure}[tb]
\centerline{\includegraphics*[scale=0.41]{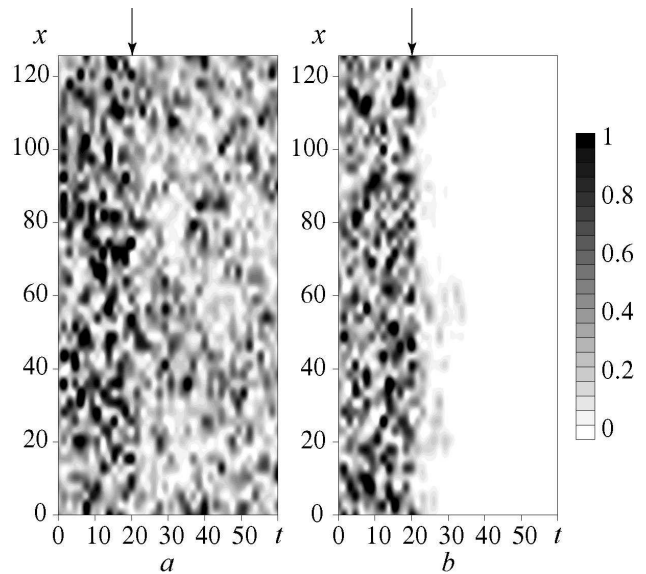}} \caption{The
dependence of the module of the difference between the states of
the response and auxiliary systems $|u(x,t)-u_a(x,t)|$ for cases
of absence ({\it a}) and presence ({\it b}) of the generalized
synchronization on time $t$ and space $x$. The coupling parameter
between the drive and response systems has been selected as
$\varepsilon=0.4$ in Fig.~({\it a}) and $\varepsilon=0.9$ in
Fig.~({\it b}). The time moments marked by arrows correspond to
the coupling switching-on between the drive and response systems
\label{fgr:SpTimeDynamics}}
\end{figure}

With the growth of the coupling strength $\varepsilon$ the
generalized synchronization between considered systems arises. The
value of the coupling strength corresponding to the onset of the
generalized synchronization is
$\varepsilon=\varepsilon_{GS}\approx 0.75$. We detected the
presence of the generalized synchronization between
unidirectionally CGLE's with the help of the auxiliary system
approach \cite{Rulkov:1996_AuxiliarySystem}. As the auxiliary
system $u_a(x,t)$ we consider the media describing by GLE
(\ref{eq:response}) which is identical to the response system
$u(x,t)$ but starts with the other initial spatial distribution,
i.e., ${u_a}(x,t_0)\neq {u}(x, t_0)$.
Fig.~\ref{fgr:SpTimeDynamics} shows the spatiotemporal
distributions of the module of the difference between the states
of the response and auxiliary systems $|u(x,t)-u_a(x,t)|$ for
cases of the absence (Fig.~\ref{fgr:SpTimeDynamics}{\it a}, small
value of coupling strength $\varepsilon<\varepsilon_{GS}$) and the
presence (Fig.~\ref{fgr:SpTimeDynamics}{\it b}, value of coupling
strength $\varepsilon>\varepsilon_{GS}$) of the generalized
synchronization regime. In this figure one can see, that in the
second case the difference of the states of the response and
axillary systems in every point of space tends to be zero after
coupling begins, which means the presence of the generalized
synchronization between the drive and response CGLE's.

To explain the mechanism of the generalized synchronization
arising, following work \cite {Aeh:2005_GS:ModifiedSystem}, we
consider the dynamics of the response system (\ref{eq:response})
as the non-autonomous dynamics of a modified spatially extended
system
\begin{equation}
  \frac{\partial u_m}{\partial t}=
u_m-(1-i\alpha_r)|u_m|^2u_m+ \nonumber
\end{equation}
\begin{equation}
 +(1+i\beta_r)\Delta u_m - \varepsilon u_m, \quad
 u_m\in[0,L],
 \label{eq:modified}
\end{equation}
under the external force $(\varepsilon v)$. Note, that the term
$-\varepsilon u_m$ brings the additional dissipation into the
modified GLE (\ref{eq:modified}).

\begin{figure}[tb]
\centerline{\includegraphics*[scale=0.5]{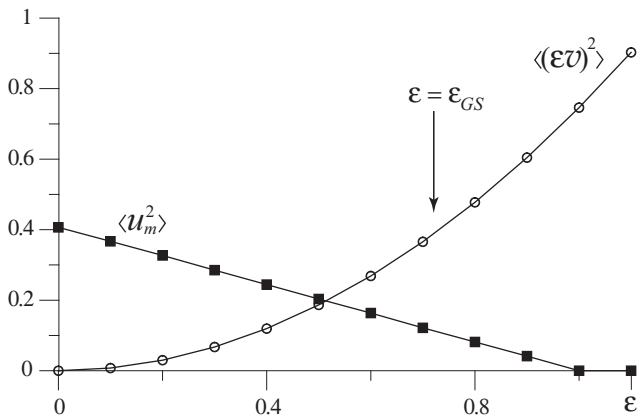}} \caption{The
dependence of averaged power of the oscillations $\langle
u_m^2(x,t)\rangle$ in the modified system ($\blacksquare$) and the
external force amplitude $\langle(\varepsilon v(x,t))^2\rangle$
($\bigcirc$) on coupling strength $\varepsilon$. The value of the
parameter $\varepsilon_{GS}$ corresponding to the onset of the
generalized synchronization regime is shown by an arrow
\label{fgr:LevelGS}}
\end{figure}

So, the control parameter $\varepsilon$ increase may be considered
as a result of two cooperative processes taking place
simultaneously. The first of them is an increase of the amplitude
of the external signal on  the response system and the second one
is the growth of the dissipation in the modified spatially
extended system~(\ref{eq:modified}).
As a result of the second process, in the modified system a
decrease of the amplitude of chaotic oscillations is observed. At
the coupling strength $\varepsilon=\varepsilon _{0}=1$ in the
spatially extended system the stable homogenous spatiotemporal
state is established rigidly in space and time.
In Fig~\ref{fgr:LevelGS}, the dependence of the square of the
amplitude of oscillations $\langle u_m^2(x,t)\rangle$ of the
modified GLE~(\ref{eq:modified}) averaged over space and time on
the parameter $\varepsilon$ is shown (symbols $\blacksquare$). One
can easily see, that the averaged amplitude of oscillations
decreases linearly with the growth of the dissipation term
$-\varepsilon u_m$ (i.e., with the increase of the coupling
strength $\epsilon$).

In work \cite{Aeh:2005_GS:ModifiedSystem} it has been shown, that
there are two mechanisms of the generalized synchroniza\-tion
arising. The first of them is determined by introducing of the
additional dissipation in the response system by means of the
dissipative term $(-\varepsilon u_m)$. If the generalized
synchronization is observed in (\ref{eq:drive}) and
(\ref{eq:response}) the modified system displays the periodic
oscillations and may undergo transition to the stable homogenous
spatiotemporal state with the growth of the parameter
$\varepsilon$ (see \cite{Aeh:2005_GS:ModifiedSystem}). This
mechanism of the generalized synchronization arising is realized
in the considered spatially extended system at
$\varepsilon=\varepsilon_0=1$.
Note, that for the considered systems the generalized
synchronization regime reveals at value of the coupling strength
$\varepsilon_{GS}$ (it is marked by an arrow in
Fig.~\ref{fgr:LevelGS}) which is less than the value of the
coupling strength $\varepsilon_{0}$ at which the stable
spatiotemporal state is established (i.e.
$\varepsilon_{GS}<\varepsilon_{0}$).

Such behaviour is detemined by the second mechanism of the
generalized synchronization arising
\cite{Aeh:2005_GS:ModifiedSystem}. Let us consider the dependence
of the square of the external force amplitude $\langle(\varepsilon
v^2\rangle)$ averaged over space and time, influenced on the
response GLE from the drive system. This dependence is shown in
Fig.~\ref{fgr:LevelGS}, (symbols $\bigcirc$).
Fig.~\ref{fgr:LevelGS} shows that the power of the drive signal
effecting  on the response system increases rapidly with the
coupling parameter growth. As a result, for
$\varepsilon=\varepsilon_{GS}$ the power of the external force
exceeds the level of own oscillations of the response system
approximately in 3 times.
It is clear, that in this case the great external force moves the
spatiotemporal state of the response system into the regions of
the phase space with the strong dissipation. So, own
spatiotemporal chaotic dynamics of the modified system (modified
GLE) appears to be suppressed and the generalized synchronization
is observed for $\varepsilon_{GS}<\varepsilon_{0}$.
It is important to note, that in the range of the coupling
strength $\varepsilon\in(\varepsilon_{GS},\varepsilon_0)$ the
generalized synchronization arising is caused by the simultaneous
action of two mechanisms, each of them brings the contribution to
the establishment of the synchronous regime.

\begin{figure}[tb]
\centerline{\includegraphics*[scale=0.5]{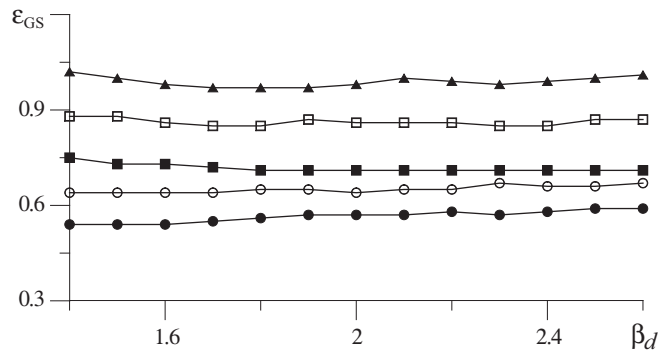}} \caption{The
dependence of the coupling strength $\varepsilon_{GS}$
corresponding to the onset of the generalized synchronization
regime on the drive system control parameter $\beta_d$ for the
different values of the control parameters
${\alpha_r=\beta_r=3.0}$ ($\bullet$), $3.5$ ($\circ$), $4.0$
($\blacksquare$), $5.0$ ($\square$), $6.0$ ($\blacktriangle$) of
the response system \label{fgr:LevelGSonBeta}}
\end{figure}

So, in the considered spatially extended system the generalized
synchronization arising is determined by two mechanisms taking
place simultaneously which causes the suppression of own chaotic
spatiotemporal oscillations by means of the additional dissipation
introducing in the spatially extended active system. In the case
of unidirectionally coupled CGLE the arising of the generalized
synchronization regime is caused by the following mechanisms.
Firstly, there is the additional dissipative terms which results
in a decrease of the magnitude of own oscillations in the response
spatially extended active system. Secondly, we observed that the
great external signal destroys completely own dynamics of the
response system and its phase state is moved into the regions of
the phase space with the strong dissipation. At the same time the
simultaneous decrease of the amplitude of own oscillations takes
place due to the first mechanism discussed above.

The last means, that the coupling strength $\varepsilon_{GS}$
corresponding to the onset of the generalized synchronization
regime in the spatially extended chaotic systems should not depend
strongly on the parameters of the drive system and (first of all)
should be defined  by the properties of the modified response
system. This statement is illustrated in
Fig.~\ref{fgr:LevelGSonBeta}, where the dependence of the coupling
strength $\varepsilon_{GS}$ corresponding to the onset of the
generalized synchronization regime on the drive system control
parameter $\beta_d$ and for the different values of the control
parameters $\alpha_r$ and $\beta_r$ of the response system is
shown. One can see that changing the drive system parameters does
not effect practically on the threshold $\varepsilon_{GS}$ of the
generalized synchronization arising in the response system with
growth of the coupling strength $\epsilon$. It confirms the
consideration of the mechanisms of the generalized synchronization
regime arising in the coupled CGLE's.

In conclusion, we have explained the generalized synchronization
arising in the unidirec\-tional\-ly coupled CGLE's and shown that
the generalized synchronization in spatially extended chaotic
systems is determined by the additional dissipation introduced
into the response system. In this case the coupling parameter
increase is equivalent to the simultaneous growth of the
dissipation and the amplitude of the external signal.

This work has been supported by U.S.~Civilian Research \&
Development Foundation for the Independent States of the Former
Soviet Union (CRDF, grant {REC--006}) and Russian Foundation of
Basic Research (grant 05--02--16273). We also thank ``Dynastiya''
Foundation.


\end{document}